# Forming Long-range Order of Semiconducting Polymers through Liquid-phase Directional Molecular Assemblies

*Minh Nhat Pham[1], Chun-Jen Su[2], Yu-Ching Huang[3], Kun-Ta Lin[1], Ting-Yu Huang[1], Yu-Ying Lai[1], Chen-An Wang[2], Yong-Kang Liaw[1], Ting-Han Lin[1], U-Ser Jeng[2], Jrjeng Ruan[1], Chan Luo[4], Ye Huang[5], Guillermo C. Bazan[6], and Ben B. Y. Hsu[1\*]*

[1]Department of Materials Science and Engineering, National Cheng Kung University, Taiwan.
[2]National Synchrotron Radiation Research Center, Hsinchu 30077, Taiwan.
[3]Department of Materials Engineering, Ming Chi University of Technology, New Taipei City, Taiwan.
[4]Intel Corporation, Hillsboro, Oregon 97124, United States.
[5]The Dow Chemical Company, Houston, TX 77077, United States.
[6]Institute for Functional Intelligent Materials, National University of Singapore, 117544, Singapore.

**ABSTRACT**

Intermolecular interactions are crucial in determining the morphology of solution-processed semiconducting polymer thin films. However, these random interactions often lead to disordered or short-range ordered structures. Achieving long-range order in these films has been a challenge due to limited control over microscopic interactions in current techniques. Here, we present a molecular-level methodology that leverages spatial matching of intermolecular dynamics among solutes, solvents, and substrates to induce directional molecular assembly in weakly bonded polymers. Within the optimized dynamic scale of 2.5 Å between polymer side chains and self-assembled monolayers (SAMs) on nanogrooved substrates, our approach transforms random aggregates into unidirectional fibers with a remarkable increase in the anisotropic stacking ratio from 1 to 11. The Flory-

Huggins-based molecular stacking model accurately predicts the transitioning order on various SAMs, validated by morphologic and spectroscopic observations. The enhanced structural ordering spans over 3 orders of magnitude in length, raising from the smallest 7.3 nm random crystallites to >14 μm unidirectional fibers on sub-millimeter areas. Overall, this study provides insights into the control of complex intermolecular interactions and offers enhanced molecular-level controllability in solution-based processes.

Keywords: intermolecular interactions, self-assembled monolayers, semiconducting polymers, directional molecular assembly, long-range ordering

*Corresponding author: hsubon@gmail.com

## Introduction

Semiconducting polymers have garnered significant attention in the field of organic electronics due to their attractive combination of electronic properties, mechanical flexibility/stretchability, and solution-processability for a broad range of device applications.[1–10] However, the random interactions among solvents, solutes, and substrates during solution-processing often disrupt molecular stacking, leading to disorder or local order with misaligned molecular orientations. This misalignment compromises the accuracy of device characterizations and calls for the establishment of long-range order with cohesive molecular orientation. Nevertheless, achieving macroscopically ordered structures while dealing with such microscopic disruptions presents a substantial challenge for solution-processed semiconducting polymers.

To tackle these challenges, two primary approaches have been investigated: increasing intermolecular coupling to circumvent disorders[11–13] and directionally coating polymer chains to suppress disorders.[14–22] While enhancing molecular planarity does improve electronic coupling and performance in the presence of disorders, reliable analysis and modeling still hinge on ordered structures such as directionality and periodicity to quantify electronic structures. Current directional coating, on the other hand, is rarely quantified in the molecular level. Therefore, this work concentrates on to overcome this quantification bottleneck existing in both approaches. Several directional solution processes have been extensively studied, such as blading,[14–16] dip coating,[17,18] directional crystallizing,[19,20] and sandwich coating.[21,22] These techniques manipulate the interactions among solvents, solutes, substrates, and process conditions to enhance the ordering of polymers. Despite their different mechanisms and coating dynamic lengths ranging from several nm[23,24] to several μm,[25–28] these ordered thin films typically consist of small crystallites with sizes around 6 to 17 nm.[16,17,19,21] This similarity in crystallite sizes suggests common microscopic dynamics underlying the ordering processes across nano-to-micrometer scales. Understanding the precise microscopic information, such as interactive scale and energy, can help develop more controllable dynamics. However, identifying a dynamically controllable state for molecules in the complex environment of solution processes remains a challenging task.

This study focuses on the identification and guidance of weakly bonded polymer bundles in highly soluble, high-temperature halogenated benzenes. Despite their pronounced solubility, these metastable bundles persist even at temperatures >100 ℃ and can be regulated through self-assembled monolayers (SAMs). The SAMs with polar carbonyl groups attract and guide small bundles to stack anisotropically along 1D-nanogrooved substrates. This approach, termed nano-templated SAMs (nSAMs), combines molecular and substrate guidance to achieve ordered stacking over a sub-millimeter scale. However, achieving long-range ordering requires the optimization of nSAMs-bundle interactions within a spatial scale of 2.5 Å, as interactions beyond this range destabilize the system and lead to disorders. To understand this angstrom-level sensitivity of ordering, we employ a Flory-Huggins-based molecular stacking model to depict the competitive enthalpic and entropic effects arising from chain length deviations between nSAMs and polymer side chains. The modeling analysis reveals that <2.52 Å conformation deviation yields the maximum interactive energy of -40.6 meV and a well-interdigitated angle 14°, which stabilize molecular stacking to activate intermolecular guiding by nSAMs. These findings highlight the importance of controlling intermolecular dynamics at the angstrom-level and underscore the necessity for precise conformation selectivity in stabilized molecular stacking that holds potential to access polymeric epitaxy on SAMs[29–31] and further advance organic electronics.[32,33]

## Result and Discussion

The liquid-phase molecular assembly plays a critical role in the creation of well-ordered thin films, governed by the intricate interactions among solute-solute, solute-solvent, and solute-substrate. If these interactions are left unregulated, they often lead to random aggregates and coils. Rigid aggregates and entangled coils both possess certain level of uncontrollability, which may freely grow into agglomerates, crystals, or fibers[34–39] with misaligned defects, hindering the necessary control in directional solution processes. In order to overcome this limitation and achieve controlled directional processing, it is necessary to establish a controllable intermolecular state that offers more freedom than aggregates but is less free than coils. This regulated dynamic state shall exhibit a degree of freedom between 2D aggregates and 3D coils. By manipulating intermolecular interactions, one can optimize this state and facilitate the formation of thin films with minimal disruption in solution processes.[21,22,40,41]

In this study, a novel methodology called intermolecular guidance is introduced as a means to achieve macroscopically ordered morphology in semiconducting polymers. The approach encompasses three main aspects: (1) identifying the controllable state of polymer bundle, (2) examining its guidance through nano-templated self-assembled monolayers (nSAMs), and (3) quantifying the driving forces behind intermolecular guidance. The first part of the study (Figure 1a) utilizes temperature-dependent UV-VIS absorption spectroscopy to examine the presence of the quasi-3D bundle state for poly(3-hexylthiophene) (P3HT) in dichlorobenzene (DCB) and dibromobenzene (DBB). By monitoring the electronic absorption at various temperatures, two distinct dynamic behaviors are observed, confirming the switchable bundle and aggregate states. The second part of the study (Figure 1b) focuses on the influence of processing temperature and nSAMs on bundle guidance. Finally, a modeling analysis predicts the transitioning order and interdigitated angles on a series of nSAMs with varying chain lengths. The predictions are supported by multiple spectroscopic and morphologic characterizations. Detailed examination and discussion of the three aspects are provided in the subsequent sections.

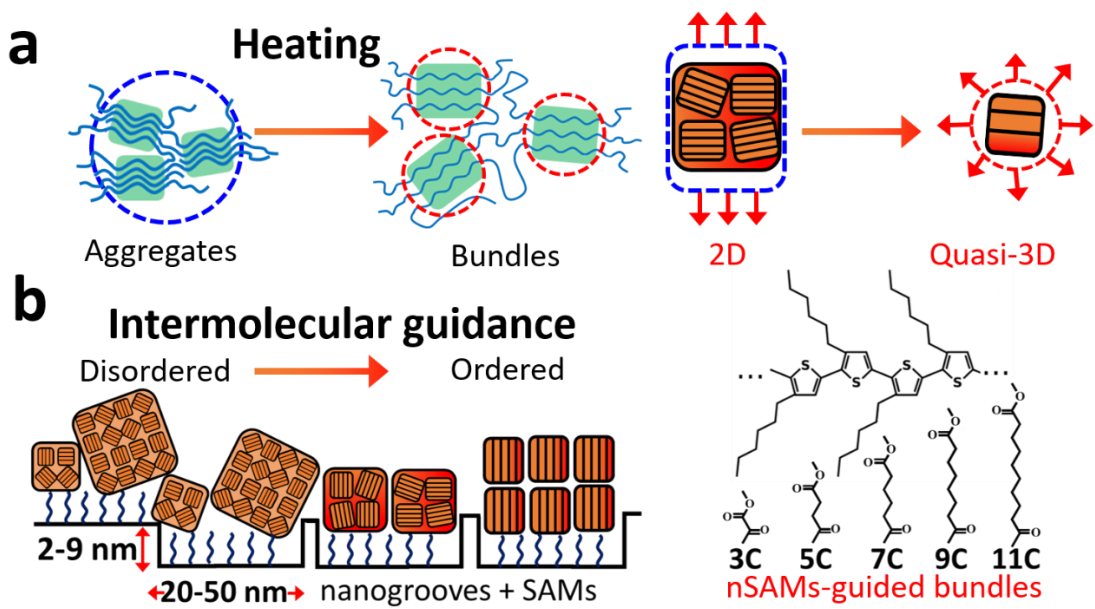

**Figure 1. (a)** DCB and DBB at rising temperature gradually enhance the solubility of P3HT aggregates (lamella) till dissolving into weakly bonded segments (bundles) that show different dynamic behaviors. (**b**) After aggregates dissolve, the optimized intermolecular interactions on the nSAMs can effectively guide the small bundles and extend their stacking order along nanogrooves to form unidirectional fibers.

**The bundle state and the thermoelectronic absorption model:** The bundle state of P3HT in solutions was examined through temperature-dependent UV-VIS absorption spectroscopy. Two solutions were prepared, each containing 0.025 wt% P3HT in either DCB or DBB. The solutions underwent a heating cycle from room temperature (RT) to 160

°C and subsequently cooled back to RT, with absorption spectra collected at 10 °C intervals. The obtained absorption spectra (Figure 2ab) show a continuous blue-shift during heating (red curves), indicating thermal expansion, and a red-shift during cooling (blue curves), indicating thermal contraction. These spectral shifts provide evidence of coupled electronic and structural transitions, corresponding to thermal strain/stress between P3HT molecules. Further analyses using small-angle and wide-angle X-ray scattering of Figure S1 confirmed the presence of 6.6 nm aggregates packed by thiophene dimers with multiple π-intervals in dilute DCB solution at RT. The X-ray scattering features and packing structures can be found in the discussion about Figure S1. In Figure 2ab, the absent intramolecular features (typically 575, 610, and 685 nm)[35,37] suggest "soft" P3HT aggregates above RT, instead of the commonly expected rigid aggregates. This supports the predominant role of intermolecular couplings between thiophene dimers in controlling spectral shifts when dimers are dynamically expanding and contracting.

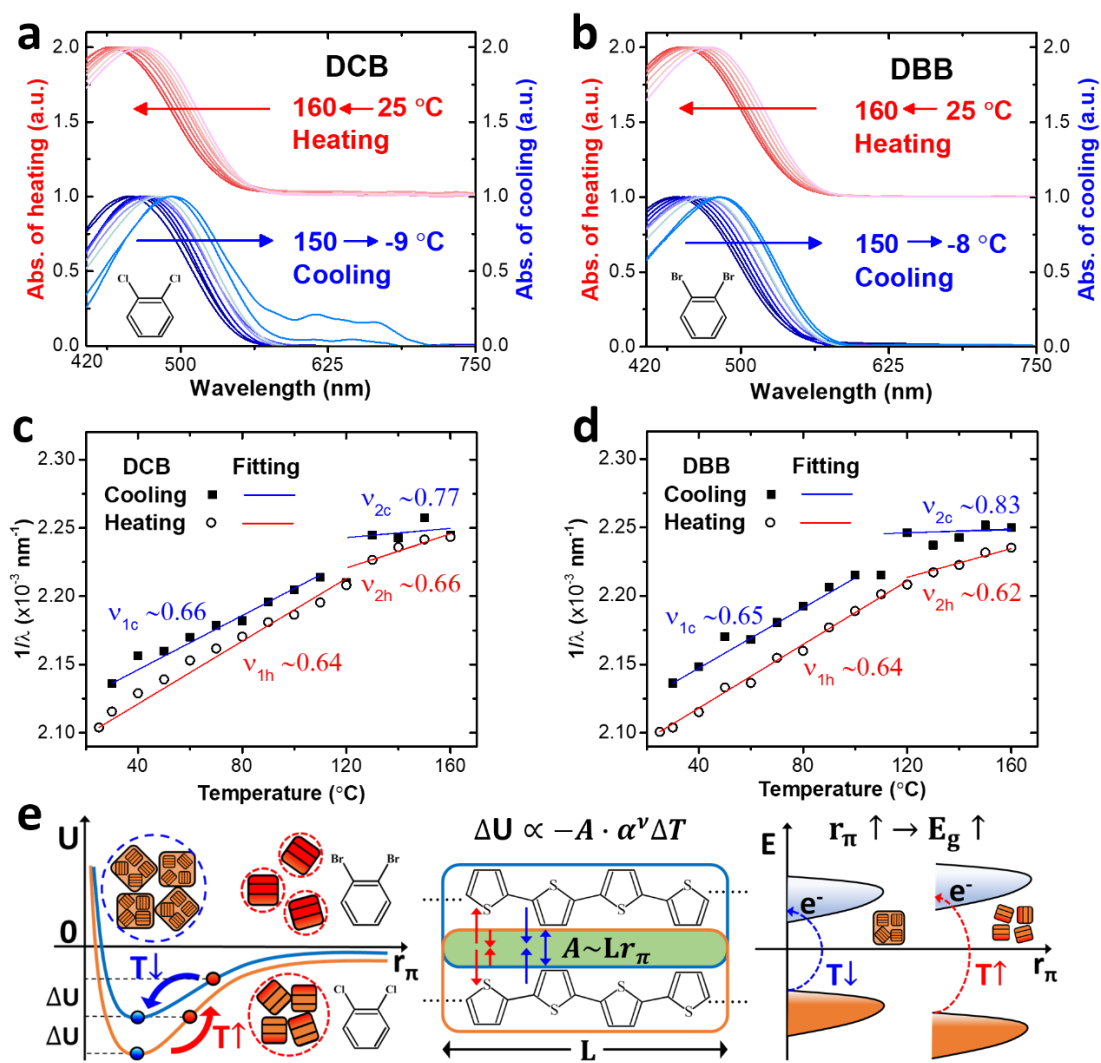

**Figure 2.** Control over the transition between P3HT aggregates and bundles in solutions. Temperature-dependent absorption spectra of the P3HT aggregates in **(a)** DCB and **(b)** DBB solutions. Spectral features clearly blue shift while heating (Red) and red shift while cooling (Blue). The $1/\lambda$-*to-T* relation of P3HT aggregates in **(c)** DCB and **(d)** DBB solutions illustrates two distinguishable dynamic behaviors with dimensionality order $\nu$ varying from 0.62 to 0.83. **(e)** Left: The coupling potentials of P3HT in DCB (light orange) and DBB (dark cyan). The deeper potential, the lower solubility, the higher dissolving temperature. Middle: The green area depicts the thermal strain that increases/decreases

intermolecular coupling in thiophene dimers upon thermal contraction/expansion. Right: The red/blue-shift electronic transitions corresponding to the thermal strain in cooling/heating.

To study the intermolecular dynamics of P3HT dimers at varying temperatures, a series of peak wavelengths were analyzed using the $1/\lambda$-to-$T$ relation depicted in Figure 2cd. The cooling curves (solid squares) exhibit two distinctive slopes separated by a plateau ranging from 100 to 120 °C, which signify a phase transition occurring in both DCB and DBB solutions. In contrast, the heating curves (open circles) for undissolved aggregates display nearly a single slope. Interestingly, in such a complex environment with vigorous solvation, both heating and cooling curves below 100 °C are almost two parallel paths as a signature of reversible electronic and structural dynamics. These linear curves therefore strongly imply that the intermolecular energetics of aggregates are governed by a single thermal expansion coefficient through thermal strain. To provide a theoretical framework for the observed spectral shifts, we devised a thermoelectronic absorption model which incorporates a dimension-dependent thermal expansion coefficient $\alpha^v$ to mediate the thermal strain and its stress-induced energy changes, as described by Equation (1). For a comprehensive analysis and calculation details, please refer to the Appendix 1 in Experimental Methods.

$$1240 \left( \frac{1}{\lambda_o} - \frac{1}{\lambda} \right) = -\beta \cdot A \alpha^v \Delta T \tag{1}$$

; $\lambda_o$ and $\lambda$ represent the peak wavelengths between the reference (subscripted as 0) and other temperatures. The thermally strained area $A = Lr_\pi$, with larger area resulting in stronger coupling, responds to temperature change $\Delta T = T - T_o$ via the coefficient $\alpha^\nu$. The values of $L$, $r_\pi$, $\alpha$, $\nu$ correspond to the intramolecular coupling length, dimer interval, thermal expansion coefficient of P3HT under isobaric condition, and its dimensionality order, respectively. Consequently, the coupling energy U within the 2D aggregates linearly shifts with the thermal strain $A\alpha^\nu\Delta T$. To quantify the observed spectral shifts, we introduce the energy density $\beta = \Delta E_{1,2}/A_o$, where $\Delta E_{1,2}$ is the total energy shifts on area $A_o = Lr_{\pi o}$ in the 1$^{st}$ and 2$^{nd}$ fitting regimes at low and high temperatures ($r_{\pi o}$ as the dimer interval at the reference temperature). Multiplying the energy density $\beta$ by thermal strain yields $\beta A\alpha^\nu \Delta T$, which can be understood as the thermal stress-induced energy change $\Delta U$ shown in Figure 2e (left). The green strained area in Figure 2e (middle) then leads to the observed spectral shifts in Figure 2e (right). Thus, Equation (1) can be further converted into Equation (2) to provide a more explicit fitting of the $1/\lambda$-to-$T$ relation.

$$\frac{1}{\lambda} = \left(\frac{\Delta E \cdot r_\pi}{1240 \cdot r_{\pi 0}} \alpha^\nu\right) \Delta T + \frac{1}{\lambda_o} \qquad (2)$$

To ensure accurate fitting of the linear spectral shifts, we adopted a theory-guided approach to determine the values of $r_{\pi 0}$ and $r_\pi$. The selection of $r_{\pi 0}$ and $r_\pi$ in both low and high temperature regimes was based on the intervals at the potentials of equilibrium, $\Delta E_2$ below zero, and zero for the low- and high-energy dimers in the theoretical work.[42] In the 1$^{st}$ regime with $T<110$ °C, $r_{\pi 0}=3.9$ Å at RT, $r_\pi=6.1$ Å at 110 °C were utilized, with $\Delta E_1= 130$ and 134 meV for DCB and DBB solutions (Refer $\Delta E$ to Figure 2cd). These $\Delta E_1$ and $r_\pi$ values before the transition plateaus are in line with the theoretical calculation for a series of cofacially-oriented thiophene dimers, where coupling potentials vary from 57 to 135 meV with $r_{\pi0}$=3.6 to 5.6 Å.[42] In the 2$^{nd}$ regime with $T$>120 °C, $r_{\pi0}$=6.4 Å at 120 °C and $r_\pi$=8.0 Å at 160 °C were chosen, with averaged energy differences $\Delta E_2$=43 and 33 meV for DCB and DBB solutions. Importantly, these theoretically determined $r_\pi$ before dissolving were also validated by the WAXS data showing the π-interval range from 2.8 to 7.2 Å. After incorporating isotropic thermal expansion coefficient $\alpha \approx 3\alpha_{1D}$=4.26×10$^{-4}$ 1/K from the 1D thermal expansion coefficient $\alpha_{1D}$=1.42×10$^{-4}$ 1/K of melted P3HT at $T$=145~200 °C,[43] we successfully extracted the dimensionality order $\nu$ corresponding to the structural dynamics of P3HT aggregates, as shown in Figure 2cd.

Investigating dimensionalities of P3HT aggregates is crucial for understanding their dynamic behaviors and potential ordering capabilities. The dimensionality values $\nu$ were fitted by the least-square method, which yields $\nu_{1h}$ and $\nu_{1c}$ ranging from 0.64 to 0.66; here, the subscript 1h/1c denotes heating/cooling in the 1$^{st}$ regime (<100 °C). These values indicate that the aggregates (lamella) behave as a 2D system ($\nu$=2/3). As the temperature increases and P3HT aggregates dissolve, $\nu$ shall approach 1.0, behaving as random, isotropic motions similar to melted P3HT. This theoretical prediction suggests a transition from a 2D behavior at lower temperatures to a 3D behavior at higher temperatures. Moving on to the 2$^{nd}$ regime (>120 °C), where thermal noise is evident, the $\nu_{2c}$ values extracted from the DCB and DBB solutions range from 0.71 to 0.85 (average 0.77) and from 0.71 to 0.97 (average 0.83), respectively; the variation of $\nu_{2c}$ is associated with the energy uncertainty of $1/\lambda_o$ varied from 3 to 18 meV ($\sim \frac{1}{2}kT$ at 160 °C) without invalidating the fitted dimensionality ($\nu>1$). These $\nu_{2c}$ values signify a higher degree of freedom for bundles compared to aggregates while still exhibiting lower freedom than random coils. Thus, the quasi-3D dynamic behavior in this regime aligns with the requirement for intermolecular guidance; high order at high processing temperature is expected. Importantly, the larger $\nu_{2c}$ values in the DBB solution compared to the DCB solution indicate that the DBB solution has higher solubility and weaker aggregation than the DCB solution. This observation is consistent with the tested solubility of P3HT, 16.3 mg/mL in DCB and 82.0 mg/mL in DBB. Overall, these findings offer valuable insights into the thermally controllable intermolecular dynamics of P3HT aggregates and serve as a foundation of guiding bundles.

Temperature-dependent UV-VIS absorption spectroscopy in conjunction with appropriate modeling has been established as an invaluable tool for investigating the structural and electronic transitions of small polymer aggregates/bundles across a broad temperature range. This technique is particularly important for studying liquid-phase dynamics, as signal of traditional X-ray scattering methods is severely degraded due to intense scattering from halogenated solvents. Although synchrotron radiation-based X-ray techniques can enhance signal strength in such an extremely dilute condition (as first demonstrated in this work), their application is restricted due to limited beam time and high-cost instrumentation. In contrast, temperature-dependent UV-VIS absorption spectroscopy is readily available on a daily basis, and provides a feasible solution to overcome these limitations by enabling the monitoring of microscopic dynamics in polymer nanoaggregates across a wide temperature range.

**Intermolecular guidance of bundles and nSAMs:** The primary objective here is to quantitatively evaluate the impact of intermolecular guidance on molecular ordering. To accomplish this with a minimal complexity, a steady and directional coating mechanism in molecular level is required. The sandwich coating method merits the directional solution flow driven by the capillary forces on the glass spacers, constantly circulating within the space between two nSAMs-fabricated substrates to align intermolecular interactions between P3HT and nSAMs.[21,22] A series of nSAMs with increasing chain lengths were introduced. The chain lengths varied from 3.9 to 13.9 Å, with increments of 2.5 Å, equivalent to two carbon-carbon bond lengths. The nSAMs, denoted as 3C, 5C, 7C, 9C, and 11C-nSAMs, were then coated with P3HT using the sandwich method. To accurately assess the ordering efficacy of the P3HT layers, the disordered top layers were removed, exposing the ordered/disordered bottom layers on the nSAMs (the peeling method was employed as described in Experimental Methods). Polarized Raman scattering was then performed to obtain anisotropic and isotropic scattering spectra of P3HT (Figure S2ab). Analyzing the ratios of anisotropic scattering intensities ($I_\parallel/I_\perp$) for the 1445 cm$^{-1}$ double-bond feature along the directions parallel and perpendicular to the nanogrooves revealed insights into the molecular orientations. Consequently, $I_\parallel/I_\perp$ of anisotropic Raman scattering corresponded to the ordering of molecules, with higher values reflecting greater molecular order. Notably, the increase in $I_\parallel/I_\perp$ in Figure 3a occurred primarily at high temperatures during the plateaus of the blue-shifting absorption, indicating that intermolecular guidance enhances the ordering of bundles rather than aggregates. Furthermore, highly ordered molecular stacking was observed exclusively on 7C-nSAMs (Figure 3b), with the highest average (max) $I_\parallel/I_\perp$ ratios of 10.8 (16.3) and 14.5 (17.5) obtained from 100 °C DBB and 160 °C DCB solutions, respectively. Consequently, the most effective intermolecular scale was found to lie within a narrow range between 6.4 Å of 5C-nSAMs and 11.4 Å of 9C-nSAMs. Within this 5.0 Å scale, intermolecular guidance via 7C-SAMs demonstrated the ability to thermally enhance the ordering of bundles with different $v_{2c}$. As depicted in Figure 3b, the ordering of DBB-processed bundles with larger $v_{2c}$ increased and reached saturation after reaching 80 °C, while the ordering of DCB-processed bundles with lower $v_{2c}$ continues to increase with temperature until 160 °C. These distinct trends are in accordance with their solubilities, as higher temperatures are required to untangle low-solubility aggregates. Consequently, the highest $I_\parallel/I_\perp$ values were observed at 160 °C and 100 °C for DCB and DBB solutions.

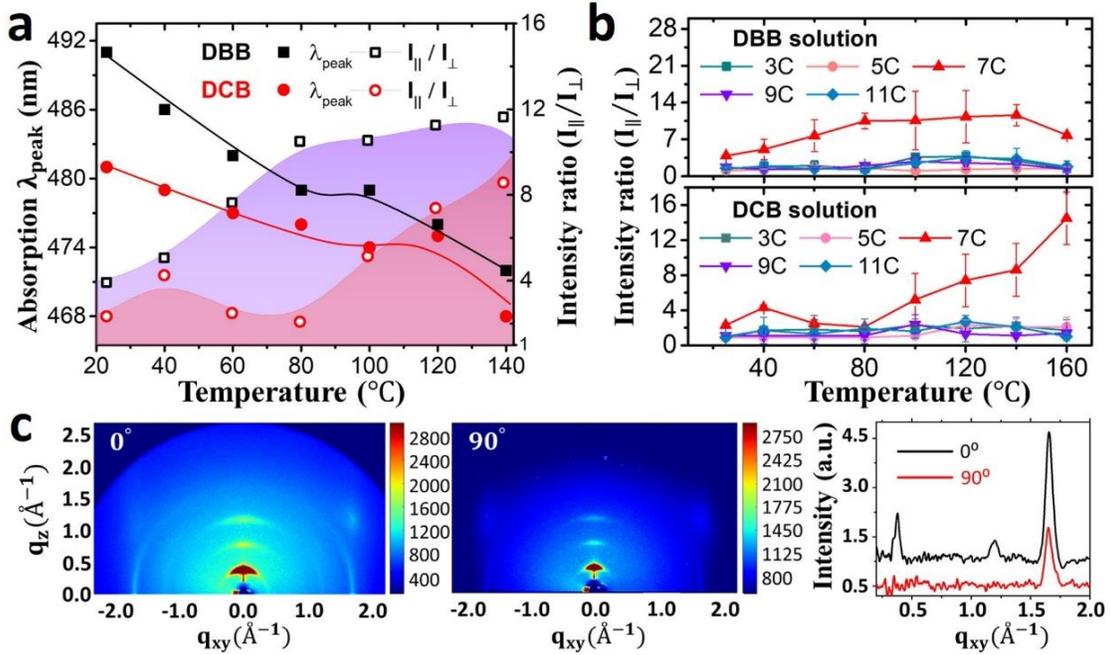

**Figure 3.** Ordering P3HT bundles by intermolecular guidance. **(a)** UV-VIS absorption peaks continuously blue-shift with increasing $I_\parallel/I_\perp$ ratios of the 1445 cm$^{-1}$ Raman scattering feature at rising temperature. **(b)** $I_\parallel/I_\perp$ of the DCB and DBB-processed P3HT nanofibers show effective directional assembly on 7C-nSAMs. **(c)** Anisotropic GIWAXS spectra parallel (0°) and perpendicular (90°) to nanogrooves on a 500×100 μm$^2$ beam area.

The alignment of P3HT bundles on different nSAMs was further examined through the orientational order parameter of P3HT chains. The orientational angle distribution, denoted as $\theta_{2D}$ was determined based on the $I_\parallel/I_\perp$ ratios of parallel and perpendicular intensities, as described by Equation (3).[44]

$$\frac{I_\parallel/I_\perp - 1}{I_\parallel/I_\perp + 1} = 2\langle cos^2\theta_{2D}\rangle - 1 \qquad (3)$$

The $\theta_{2D}$ values of $\langle cos^2\theta_{2D}\rangle$, which represent the averaged orientational distribution of the P3HT chains relative to the nanogroove direction ($\theta_{2D}=0°$), were calculated from the equation. It was found that the 100°C-DBB-processed P3HT on 7C-nSAMs exhibited a narrower distribution of 17.1° compared to the other nSAMs, indicating a more cohesive alignment on 7C-nSAMs. The same trend of $\theta_{2D}$ was observed from the 160°C-DCB-processed P3HT. These findings, as shown in Table 1, demonstrate that even in the presence of aggressive solvents and thermal fluctuation, optimized intermolecular interactions can still stabilize molecular stacking and direct the alignment of tens-nm long polymer chains along a specific direction.

**Table 1.** The structural and order parameters extracted from the AFM topographies and polarized Raman scattering spectra of the processed P3HT films on different nSAMs.

| nSAMs | 3C | 5C | 7C | 9C | 11C |
| --- | --- | --- | --- | --- | --- |
| Chain length of nSAMs (Å)[a] | 3.9 | 6.4 | 8.9 | 11.4 | 13.9 |
| Intensity ratio $(I_\parallel/I_\perp)$[b] | 2.2 | 2.0 | 10.8 | 2.2 | 2.1 |
| Intensity ratio $(I_\parallel/I_\perp)$[c] | 1.7 | 2.1 | 14.5 | 1.4 | 1.0 |
| Angle distribution $(\theta_{2D})$[b] | 34.0 | 35.3 | 17.1 | 34.0 | 34.6 |
| Angle distribution $(\theta_{2D})$[c] | 37.5 | 34.6 | 14.7 | 40.2 | 45.0 |
| AFM roughness (nm)[b] | 3.6 | 1.5 | 1.1 | 1.5 | 2.3 |
| Fiber width (nm)[b] | 90.5 | 65.2 | 36.5 | 61.2 | 56.7 |

[a]These values are obtained by Avogadro software. [b]DBB-processed P3HT at 100 °C. [c]DCB-processed P3HT at 160 °C.

In order to investigate the extent of molecular ordering, grazing incidence wide angle X-ray scattering (GIWAXS) was utilized to analyze the as-cast P3HT on 7C-nSAMs. Despite the presence of disorders in the as-cast film, the anisotropic scattering patterns displayed in Figure 3c demonstrate the existence of extended molecular order over a large area of sub-millimeter beam spot (500 μm × 100 μm). Notably, this region is considerably larger than the 3 μm-diameter Raman laser spot. The sharp π-π scattering peaks observed in Figure 3c (right) exhibit a rarely observed narrow full width half maximum (FWHM) $\Delta q$=0.069 Å$^{-1}$, which confirms the presence of the smallest crystallites measuring 7.3 nm in both top disordered and bottom ordered layers. The combined outcomes of anisotropic Raman and X-Ray scattering provide solid evidence that anisotropic molecular stacking of $I_\parallel/I_\perp$>10 can be attributed to the alignment of 7.3 nm P3HT crystallites in the bottom layers. Conversely, the top layers are characterized by randomly oriented crystallites. These polymer chains display a strong affinity towards 7C-nSAMs and preferentially align with the fiber direction, which is contrary to the commonly occurred chain folding perpendicular to the fiber direction.[45–48] This suggests that intermolecular guidance serves as a compelling dynamic process to counteract chain folding and thermal fluctuations, thereby facilitating macroscopic order along the designated direction in hot solutions.

Next, we utilized atomic force microscope (AFM) and transmission electron microscope (TEM) to study the morphologic evolution of P3HT crystallites on different nSAMs under identical processing conditions. Surprisingly, AFM images in Figure 4a showed a disorder-order-disorder transition of long fibrotic structures from 3C to 11C-nSAMs. The roughness and fiber width, analyzed using NanoScope software, exhibited a high-low-high trend corresponding to a low-high-low morphological and molecular orders (See $I_\parallel/I_\perp$ in Table 1). The microscopic and macroscopic orders were simultaneously influenced by the chain length of the nSAMs. Increasing the chain length from 3C to 7C-nSAMs led to enhanced cohesiveness and the transformation of twisted, granular fibers to smooth, unidirectional fibers. However, further increases in chain length resulted in the wiggling and voiding of aligned fibers on 9C-nSAMs, followed by randomized fibers on 11C-nSAMs. Both shorter and longer chain lengths than 7C-nSAMs create a higher level of randomness at the bulk scale. The ordered P3HT fibers on 7C-nSAMs exhibited the smallest fiber width, least roughness, and the most cohesive orientation angle, highlighting the significance of matching conformations between side chains and nSAMs in forming compact and directional fibrotic structures.

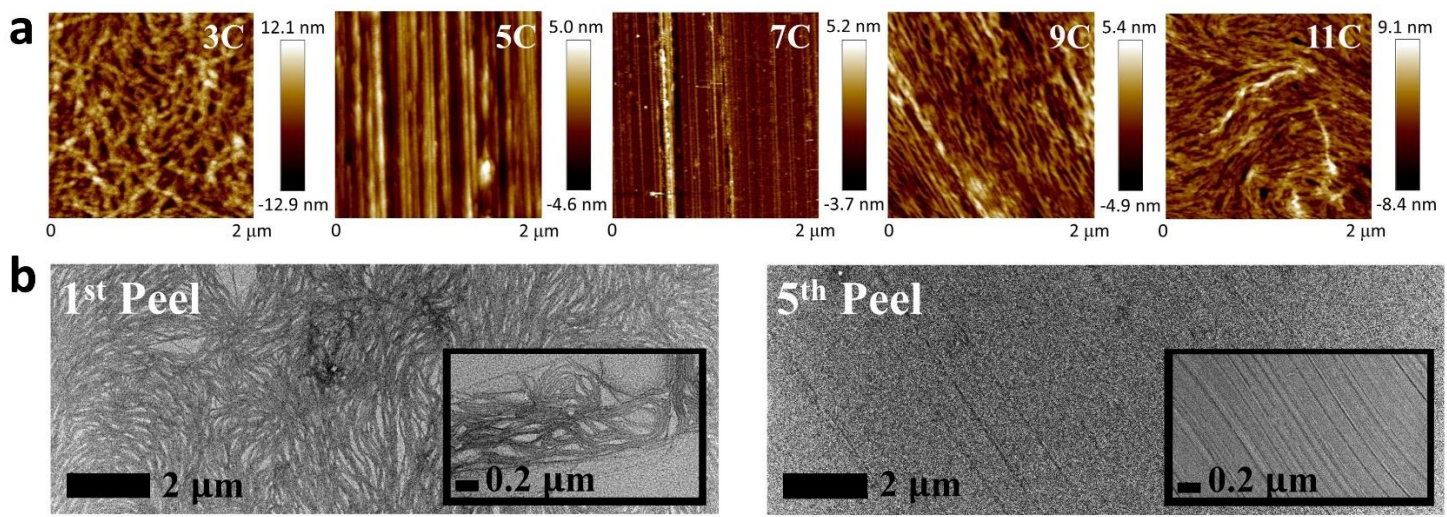

**Figure 4.** Morphologic transitions of the P3HT on various nSAMs processed from DBB solutions at 100 ℃. **(a)** A disorder-order-disorder transition is observed from 3C, 5C, 7C,

9C, to 11C-nSAMs. 7C-nSAMs with the strongest guiding effect demonstrates the highest anisotropic structures and the lowest morphologic variations. **(b)** The zoom-in TEM images of the top and bottom P3HT layers. Opposite to the top disordered fibers, the bottom layers approaching to 7C-nSAMs show macroscopically-aligned fibrotic structures (See Figure S4 for the full-size image).

To further investigate the depth-dependent morphological order in the 100 ℃-DBB-processed P3HT film, the peeled P3HT layers from different depth were characterized using TEM. In Figure 4b, the top layers displayed fibers without specific directionality, while the bottom layers exhibited unidirectional fibers. The deeper layer showed the higher order, consistent with the increasing molecular order ($I_\parallel/I_\perp$) with peeling times shown in Figure S3. 7C-nSAMs-guided fibers consisted of long directional fibers ~14 μm in sight of Figure S4, extended by aligning 7.3 nm crystallites over a large area. Unguided crystallites in top layers, on the other hand, formed disordered fibers. In the absence of SAMs, nanogrooves failed to align the P3HT chains in a specific manner, resulting in disordered fibers (Please see the isotropic topography and $I_\parallel/I_\perp$ in Figure S5). Additionally, the surface energies on the SAMs did not align with the expected low-high-low molecular order (Refer to the water contact angles in Figure S6 and Table S1). It was found that neither nanogrooves nor SAMs alone could align molecules in the long term. Instead, the combined use of SAMs and nanogrooves as nSAMs synergistically provided intermolecular guidance to extend SAMs-induced local order[49–52] into bulk-scale order.

Thus, achieving macroscopically ordered morphologies requires a sub-nanoscale spatial match between guiding and guided molecules, emphasizing the importance of establishing a molecular stacking model to quantitatively understand the dynamics of SAMs influence.

**The Flory-Huggins-based stacking model:** Here, we aimed to gain a deeper understanding of the molecular stacking mechanism by modeling. Specifically, our focus was on the interactions between nSAMs and a short P3HT segment, hypothesizing that the interdigitation between nSAMs and P3HT side chains plays a critical role in stabilizing the molecular stacking. To investigate this hypothesis, we calculated the interactive energy using $\Delta G_{int} = \Delta H_{int} - T\Delta S_{int}$, where $\Delta G_{int}$, $\Delta H_{int}$, $\Delta S_{int}$, and $T$ represent interactive Gibbs free energy, enthalpy, entropy, and temperature, respectively. To describe $\Delta H_{int}$ contribution resulted from intermolecular interactions, we utilized the Lennard-Jones potentials, focusing on the pairs formed between carbonyl group–thiophene ring and hexyl side chain-silicon substrate. The distance $r_{1,2}$, attraction strength $\varepsilon_{1,2}$, and radii $\sigma_{1,2}$ of the interacting molecules in the pairs were taken into account using Equation (4).

$$\Delta H_{int} = 4\varepsilon_1 \left[\left(\frac{\sigma_1}{r_1}\right)^{12} - \left(\frac{\sigma_1}{r_1}\right)^{6}\right] + 4\varepsilon_2 \left[\left(\frac{\sigma_2}{r_2}\right)^{12} - \left(\frac{\sigma_2}{r_2}\right)^{6}\right] \qquad (4)$$

; the subscript 1 and 2 correspond to the two different molecular pairs. By using the equations $d(\Delta H_{int})/d\, r_1 = 0$ and $r_2 = r_1 + d - l$ shown in Figure 5a, we determined that the equilibrium $\Delta H_{int}$ increases with the chain lengths of nSAMs; $d$ is the distance between carbonyl group and silicon surface while $l$ is the length of a hexyl side chain. All parameters used in calculations are provided in Table S2, and the resulting $\Delta H_{int}$, $r_1$, and

$r_2$ at thermal equilibrium are collected in Table S3. Our analysis revealed that the strength of the attraction increases with the chain lengths, showing the strongest attraction (-75.3 meV) on 7C-nSAMs and dropping a bit on 9C and 11C-nSAMs. This value is comparable to -74.2 meV of the π-π stacked thiophene dimer,[42] without overwhelming the π-π stacking of P3HT. Additionally, longer nSAMs will experience a greater decrease in $\Delta G_{int}$ due to an increase in entropy $\Delta S_{int}$, leading to the conformational optimization on 7C-nSAMs. Next, we introduce the entropic effect.

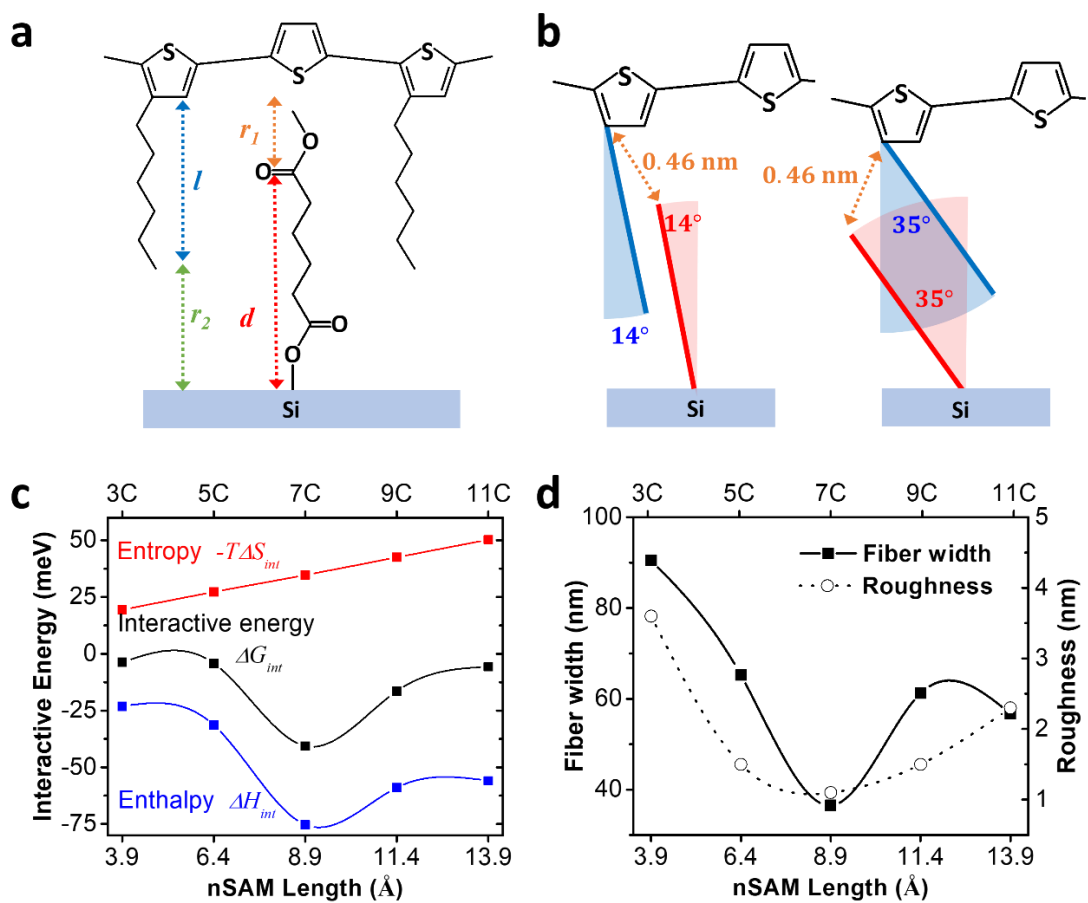

**Figure 5. (a)** The initial positions before stabilizing interdigitation. **(b)** The tilt angles 14° and 35° of the stabilized interdigitation between hexyl side chains and 7C-nSAMs. **(c)** The simulated interactive energy, using the Flory-Huggins-based stacking model, with various

length variation between SAM chains and P3HT side chains. **(d)** The fiber widths and roughness values of the processed films on different nSAMs.

To model the entropy, we allowed the interdigitated chains to tilt while keeping the strongest $\Delta H_{int}$ at the shortest $r_1$ for each nSAMs. This chain tilting can be understood as space occupancy that provides the changed entropy $\Delta S_{int}$ before and after interdigitation. The Boltzmann relation was then employed to describe $\Delta S_{int}$ of nSAMs by Equation (5).

$$\Delta S_{int} = k_B ln\left(\frac{\Omega_i}{\Omega_f}\right) \quad (5)$$

; $k_B$ is the Boltzmann constant and $\Omega_{f,i}$ are the accessible microstate numbers of all spatial configurations for the free and interdigitated nSAMs, respectively. Using the Flory-Huggins solution lattices, we quantified the configurations of nSAMs and obtained their entropies (See Figure S7 and more discussion in the supporting information). Our calculation revealed that the entropies of nSAMs increase as the chain lengths increase, compromising the enthalpic effect of longer nSAMs to maximize $|\Delta G_{int}|$ at 7C-nSAMs (See Table 2). The calculation and discussion about spatial configurations can be found in the Appendix 2 and S5 section in the SI. Detailed parameters used for calculating $\Delta S_{int}$ and $\Delta G_{int}$ are all summarized in Table S4.

**Table 2.** The calculated $\Delta H_{int}$, $\Delta S_{int}$, $\Delta G_{int}$, and tilt angles of P3HT on nSAMs.

| nSAMs | | 3C | 5C | 7C | 9C | 11C |
|---|---|---|---|---|---|---|
| Enthalpy $\Delta H_{int}$ (meV) | | -23.1 | -31.4 | -75.3 | -58.9 | -56.0 |

| Entropy $-T\Delta S_{int}$ (meV) | 19.4 | 27.2 | 34.7 | 42.5 | 50.3 |
| Free energy $\Delta G_{int}$ (meV) | -3.7 | -4.2 | -40.6 | -16.4 | -5.7 |
| Tilt angle of nSAMs (°) | 17 | 17 | 14/35[a] | 13/22[a] | 11/17[a] |

[a]Two tilt angles are energetically allowed.

In our model, the tilt angles $\alpha$ of the interdigitated chains for different nSAMs were all calculated to maximize both $|\Delta H_{int}|$ and $|\Delta S_{int}|$, giving two boundary conditions: the equilibrium packing distance $r = r_1$ and the equalized lateral separation with the two side chains ~4.0 Å seen in Figure S8. These conditions were utilized to derive Equation (6) and (7):

$$r\cos\phi + d\cos\alpha = l\cos\alpha + r_2 \qquad (6)$$

$$r\sin\phi + d\sin\alpha = 4 \qquad (7)$$

where $r$ represents carbonyl group–thiophene ring distance, which $r$ varies with different $\alpha$, defining the tilting of nSAMs chains and hexyl side chains within a range from 0° to 90°. $\phi$ is the tilt angle of the side chain used to compute normal and lateral components at the distance $r$, shown in Figure S8. Solving Equation (6) and (7) can find the $r(\alpha)$ relation described by Equation (8):

$$r = \frac{4 - d\sin\alpha}{\sin\left[\tan^{-1}\left(\frac{4 - d\sin\alpha}{l\cos\alpha + r_2 - d\cos\alpha}\right)\right]} \qquad (8)$$

Using Equation (8), we were able to solve for the angle $\alpha$ at the equilibrium distance $r(\alpha) = r_1$. The optimum $\alpha(r_1)$ values for each nSAMs were then determined (the red-highlighted values in Table S5), resulting in the tilt angles varying from 17° to 11° as the closest packing schemes (Figure S9) from 3C to 11C-nSAMs. Interestingly, for chain lengths >5C-nSAMs, two energetically-allowed interdigitated angles were observed (See Table 2), such as the 14° and 35° interdigitated angles 7C-nSAMs exhibited in Figure 5b. The angle of 14° matched the reported 13.7° of the closest packed P3HT lamella from literatures and had been experimentally verified by the TEM image and low-temperature X-ray scattering.[53,54] Meanwhile the other angle of 35° fell within the reported theoretical range of 30° to 36°.[55,56] This presence of two packing modes suggests a polymorphic state in solution-processed semiconducting polymers, even at the maximum $|\Delta G_{int}|$.

In addition, Figure 5c illustrates that the maximized interactive energy of -40.6 meV can be achieved by 8.9 Å-long 7C-nSAMs with hexyl side chains. The Gaussian fitting of the $\Delta G_{int}$ (Figure S10) yields a sub-nanoscale FWHM of 2.52 Å, consistent with the experimental observation. On the other hand, the weak enthalpy $\Delta H_{int}$ of 3C-nSAMs is insufficient to stabilize bundle stacking while 11C-nSAMs fail to properly interdigitate due to the strong entropy $-T\Delta S_{int}$ at high temperature. The fibers formed on 5C-nSAMs and 9C-nSAMs with stronger $\Delta G_{int}$, compared to those on 3C-nSAMs and 11C-nSAMs, exhibit less granular and more directional characteristics, indicating some initial signs of stabilization toward optimized ordering as aforementioned in Figure 4a. The trend of $\Delta G_{int}$ from 3.9 Å to 13.9 Å yields a high-low-high structural variation, which closely correlates with the roughness and fiber width in Figure 5d. Smaller morphologic randomness suggests that the closest packing of P3HT occurs on 7C-nSAMs.

This work demonstrates that controlled intermolecular dynamics with well interdigitated chains enhance long-range ordering by stabilizing molecular stacking in the closest packed condition, which is energetically favored within the optimized dynamic scale 2.5 Å. Figure 6 intuitively illustrates that beyond this optimized scale, both weak enthalpy and strong entropy contribute to small $|\Delta G_{int}|$ and destabilize stacking. This angstrom-level mismatch will lead to bulk-scale disorders, highlighting the importance of conformation compatibility for macroscopic ordering.

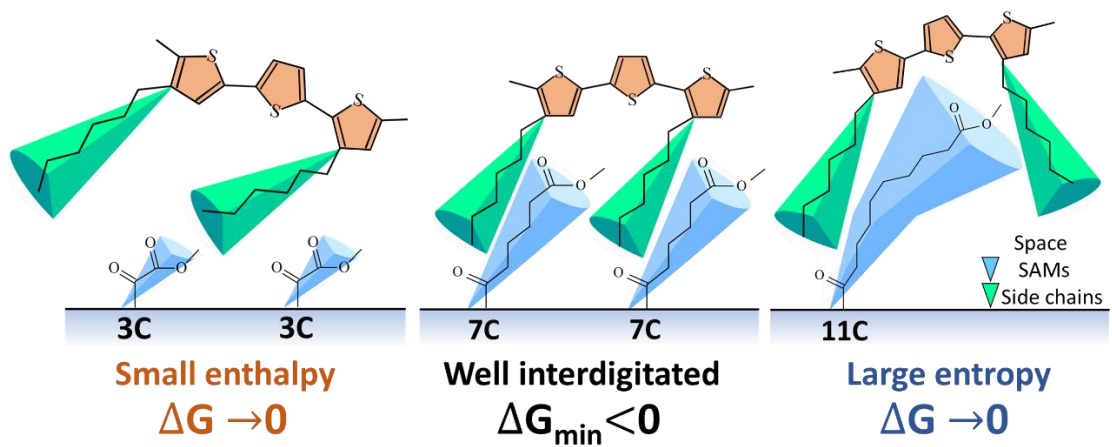

**Figure 6.** The most negative $\Delta G_{min}$ on 7C-nSAMs can stabilize intermolecular guidance via proper interdigitation while the mismatched chain lengths either degrade enthalpy or enhance entropy to increase disorders.

## Conclusions

In summary, long-range order in solution-processed semiconducting polymers can be systematically achieved by accessing a controlled dynamic state to activate its intermolecular guidance through well interdigitated nSAMs. Temperature-dependent UV-VIS absorption spectroscopy and the novel thermoelectronic model provide a reliable probing to identify the controllable bundle state with quasi-3D dynamics in solutions. The manipulations in aggregate dimensions and dynamic scales within 2.5 Å between bundles and nSAMs synergistically enhance ordering over sub-millimeter areas. Additionally, the Flory-Huggins-based molecular stacking model successfully determine the optimum interdigitation conditions such as the interactive energy, length scale, and tilt angles controlled by polymer side chains and nSAMs. By harnessing microscopic dynamics, macroscopic order can be reliably obtained. The proposed methodology here serves as a clear molecular guideline of choosing or designing the optimum conformations to effectively activate intermolecular guidance and achieve long-range ordering in liquid-phase molecular assembly.

## Experimental Methods

**Solution preparation:**

Semiconducting poly(3-hexylthiophene) (P3HT) of molecular weight $M_w$=14.4 kDa and $M_w/M_n$=1.96 was synthesized using Grignard metathesis[57] and was purified by Soxhlet extraction. Anhydrous dichlorobenzene (DCB) and dibromobenzene (DBB) were purchased from Sigma-Aldrich. The 0.025 wt% solutions of P3HT in DCB and DBB were prepared. DBB has about 5-time larger solubility for P3HT than that of DCB, 82.0 mg/mL compared to 16.2 mg/mL; the two solubilities are consistent with the reported values, 84.1 and 14.7 mg/mL in the literature.[58] They were chosen to exercise the solvation effect in the thermoelectronic absorption model.

**Temperature-dependent UV-VIS absorption of P3HT aggregates:**

Two solutions of 0.025 wt% P3HT in DCB and DBB were prepared. The size of P3HT aggregates was found ≲6.6 nm with three dominant π-intervals of 7.2, 4.1, and 2.8 Å, using small-angle and wide-angle X-ray scattering (See the discussion about Figure S1). Temperature-dependent UV-VIS absorption spectra of the two P3HT solutions were collected by a fiber-coupled holder with PID temperature control. The light source of Ocean Optics LS-1 provided a smooth spectrum in UV-VIS-NIR wavelengths from 360 nm to 2000 nm and the signal transmitted through solution cuvettes was read by a mini-spectrometer Ocean Optics USB2000+. The temperature-dependent absorption spectra in Figure 2a and 2b were referenced to the transmission spectra of pure solvents at the same temperatures and all spectra were normalized. Therefore, the observed blue/red-shifts represented the energy change per electronic unit upon temperature change; here, thiophene dimers are the smallest units in P3HT aggregates. The presence of thermal noise varying from 3 to 18 meV ($kT$=14 meV at 160 °C), was found to affect $1/\lambda_0$ values in the second regime (T>120 °C) and change the fitted values of dimensionality order ν so an averaged ν value is suggested.

**The preparation of nano-templated SAMs:**

To optimize intermolecular guidance, we prepared five self-assembled monolayers purchased from Sigma-Aldrich. Methyl adipoyl chloride is denoted as 7C while other molecules with identical functional groups, methyl acetate and chlorine, but different chain lengths are labeled as 3-carbon (3C), 5-carbon (5C), 7-carbon (7C), 9-carbon (9C), and 11-carbon (11C). Their chain lengths are estimated by Avogadro software and the optimized molecular structures come out their lengths 3.9, 6.4, 8.9, 11.4, and 13.9 Å in increments of 2.5 Å. The hexyl side chains of the P3HT are 7.7 Å, estimated by Avogadro as well. All the chemical structures are shown in Figure 1a and 1b.

The nanogrooves on silicon substrates were fabricated by the lapping films of 100 nm diamond nanoparticles.[59,60] Groove dimensions were approximately 20 to 50 nm wide and <10 nm deep as seen in the AFM image of Figure S5. These lapping films were purchased from Allied High Tech Product, 3M, and Buehler. Please note that the lapping film quality plays a crucial role in ordering and the venders may have quality variation that shows diamond aggregates of 1 to 2 μm in sizes.[61] Long-term monitoring on the grooved surface is recommended. After UV-ozone cleaning for 15 minutes, the grooved substrates were passivated respectively by the 20 mM solutions of the five different nSAMs in anhydrous toluene at room temperature (RT) for 3 hours. The increase in contact angles of the treated silicon substrates after the reactions confirmed that five SAMs were successfully anchored on the substrates, as shown in Table S1 and Figure S4. P3HT was then coated onto the passivated substrates through the sandwich process at various temperatures.

**X-Ray scattering spectroscopy:**

Examining the interactive scale and dimensionality of small P3HT aggregates is a very challenging task due to the severely-scattered weak signal of light-carbon backbones with the presence of abundant heavy halogenated atoms in very dilute solutions. A synchrotron radiation-based X-ray station is necessary. P3HT aggregates in 0.05 wt% DCB solutions were characterized by small- and wide-angle X-ray scattering (SAXS and WAXS) using at the 13A BioSAXS beamline of the Taiwan Photon Source (TPS) in National Synchrotron Radiation Research Center (NSRRC). A 20 keV X-ray beam in the transmission mode was utilized for the sample solutions sealed in a quartz capillary (with dia. of 1 mm and thickness of 20 µm). Detailed characterizations and instrumentation can be found in Section 1 of the supporting information and the relevant literatures.[62,63] Anisotropic X-ray scattering on a large area was obtained by glance-incident wide angle X-ray scattering (2D-GIWAXS) on Beamline 11-3 at Stanford Synchrotron Radiation Lightsource (SSRL).

**Imaging morphology:**

The topographic images of P3HT on different nSAMs were obtained by tapping mode scan of Bruker Innova atomic force microscopy (AFM). Additional to the topography, we peeled the thin films layer by layer to unmask the hidden morphologies of the bottom layers. A phenol novolac-based epoxy, EPO-TEK 353ND, was purchased from Epoxy Technology to peel the top disordered P3HT. For inner-layer morphology, the films were peeled through the deposited carbon films instead of novolac epoxy and the peeled layers at different depths were individually imaged by the transmission electron microscope,[64] JEOL JEM-1400 instrument operated at 120 kV. Their bright-field images were recorded via a Gatan digital camera. TEM found that ordered P3HT fibers are distributed on the bottom layers above nSAMs and disordered ones are on top of the films.

**Polarized Raman scattering:**

We employed polarized Raman scattering using a single-mode DPSS 532 nm laser. The polarization ratio ~100:1 of $TEM_{00}$ (horizontal: vertical) was further enhanced by a 45° polarizing plate beamsplitter with an extinction ratio >10000:1 (horizontal: vertical) for a linearly polarized input, purchased from ThorLABs. Meanwhile, a half-wave plate was utilized to tune the polarization and measure angle-dependent scattering charts. An Acton Princeton SP2300i spectrometer equipped with an Andor CCD camera iDus 401 collected and read the scattering signal. Prior to measuring anisotropic Raman scattering, we peeled the top disordered layers by the novolac epoxy. The molecular stacking of P3HT was obtained on an area of focused laser beam, approximately 3 μm in diameter. The anisotropic and isotropic intensity ratios $I_{\parallel}/I_{\perp}$ of the 1445 cm$^{-1}$ peaks for the ordered and disordered P3HT films are provided in Figure S2ab.

# Appendix:

## 1. The thermodynamics of thermoelectronic absorption

In order to establish a relationship between linear electronic transitions upon temperature change, it is necessary to develop a thermodynamic model that connects the structural and electronic responses. We first consider a molecular dimer that undergoes changes in volume and internal energy when subjected to heating, as shown in Equation

(A1).

$$dU = TdS - PdV \tag{A1}$$

; the internal energy $U$ represents the dimer bonding from intermolecular coupling and the thermodynamic parameters $T$, $S$, $P$, and $V$ are temperature, entropy, pressure, and volume of a dimer. To analyze its energy change upon heating and expanding, we incorporate Equation ($A1$) and the Maxwell relations to the internal energy change $dU$ with respect to $V$ and $T$ and obtain Equation ($A2$).

$$\begin{aligned} dU(V,T) &= \left(\frac{\partial U}{\partial S}\right)_V \left(\frac{\partial S}{\partial V}\right)_T dV + \left(\frac{\partial U}{\partial V}\right)_T \left(\frac{\partial V}{\partial T}\right)_P dT \\ &= T\left(\frac{\partial S}{\partial V}\right)_T dV - P\left(\frac{\partial V}{\partial T}\right)_P dT \\ &= T\left(\frac{\partial P}{\partial T}\right)_{V^g} dV - P\left(\frac{\partial V}{\partial T}\right)_P dT \end{aligned} \tag{A2}$$

; $V^g$ in the subscript of $\partial P/\partial T$ is the constant molar volume of the evaporated molecules, which satisfies the van der Waals gas model. Equation ($A2$) can be explained by considering two terms: The first term represents the heat required for the evaporated molar volume of gas from liquid, $\Delta U = \int T(\partial P/\partial T)_{V^g} dV$, which is associated with the estimated bonding energy density in the Hansen and Hildebrand theory of solubility under isovolumetric conditions.[65–67] The second term describes the behavior of a freely expandable dimer in solution at rising temperature, satisfying the isobaric condition where the internal pressure of molecules remains constant as described by Equation ($A3$).

$$d\left(\frac{U}{V}\right) = -\frac{P}{V}\left(\frac{\partial V}{\partial T}\right)_P dT = -P\left(\frac{\partial V/V}{\partial T}\right)_P dT = -P\alpha dT \tag{A3}$$

; $\alpha$ represents the isobaric thermal expansion coefficient based on a 3D-expanding volume. By considering the energy change of dimers in a 2D aggregate instead of a 3D one, we hence replace $\Delta U/V$ and $P$ with the 2D energy density $\Delta U/A$ and its surface tension $F/L$. These substitutes, sharing identical dimensions $F/L \sim d(U/A) = \Delta E/A_0$, allow us to analyze the correlation between the coupling energy and temperature of thiophene dimers in a 2D aggregate under the isobaric condition. The negative sign accounts for the decrease in coupling energy density caused by thermal expansion. After introducing a dimensionality order $v$ for $\alpha$ to account for the equipartitioning entropy, we correct the dimension inconsistency between $\alpha$ and $U/A$ (ideally, $v$=1, 2/3, and 1/3 for 3D, 2D, and 1D structures). Then Equation ($A4$) can well describe the linear spectral shifts along with structural dynamics as Equation (1) demonstrates.

$$dU = -\frac{\Delta E}{A_o} A \cdot \alpha \cdot dT \qquad (A4)$$

$$\rightarrow \Delta U = -\beta \cdot A \cdot \alpha^v \cdot \Delta T$$

; where $\Delta E$ represents the blue/red-shift energy in the fitting temperature regime and $A_o$ is the area at reference temperature. The proposed thermoelectronic absorption model clearly describes the thermodynamic relation between thermal expansion and electronic transition, indicating that the strong intermolecular coupling dictates both structural and electronic dynamics. Thus, the model can successfully acquire sub-10 nm structural dynamics through straightforward UV-VIS absorption spectroscopy, without using complicated instrumentation like X-ray scattering techniques.

## 2. The entropy and interdigitation of nSAMs

The spatial configurations before and after interdigitation are represented by $\Omega_f = Z_f{}^n$ and $\Omega_i = Z_i{}^n$, where $n$ is the number of atoms along the main chains of nSAMs, and $Z_f$ and $Z_i$ are the coordination numbers for each atom in a unit cell.[68] At the free state, nSAMs atoms have a higher degree of freedom than that of the interdigitated ones, where nSAMs are constrained between the two hexyl side chains to result in $Z_f > Z_i$. Thus Equation ($A5$) produces $-T\Delta S_{int}>0$, behaving as a counter force for the enthalpy $\Delta H_{int}$.

$$\Delta S_{int} = n \cdot k_B \ln\left(\frac{Z_i}{Z_f}\right) \qquad (A5)$$

While an atom on nSAMs is placed adjacent to the side chains, it cannot extend beyond the side chains so the next atom will have a $Z_f - 1$ coordination. Due to fast thermal motion at high temperature, the bonding coordination of nSAMs shall quickly switch between the constrained and free states. Hence, we incorporate an effective $\langle Z_i \rangle$ for Equation ($A5$).[68,69] Detailed discussion and calculation are provided in Section 5 of the supporting information.

**Supporting Information**

Small-angle and wide-angle X-ray scattering spectra; anisotropic Raman scattering intensity ratios; polarized Raman scattering spectra of ordered and disordered P3HT films; the increase of anisotropic Raman scattering intensity ratios $I_\parallel/I_\perp$ with peeling times; full-size TEM image of directional fibers; topographies of nanogrooves, amorphous P3HT, and its isotropic Raman scattering spectrum; images of water contact angles on various nSAMs; collected water contact angles on SAMs and anisotropic $I_\parallel/I_\perp$ ratios of P3HT on nSAMs; cartoons of interdigitation and the corresponding Flory-Huggins' solution lattices; parameters of the Lennard-Jones potentials used in calculating $\Delta H_{int}$(r); the calculated equilibrium $\Delta H_{int}$ at the optimum distances of $r_1$ and $r_2$; parameters of the Flory-Huggins-based stacking model used in calculating $\Delta S_{int}(r)$ and $\Delta G_{int}(r)$; schematic cartoon of the geometric relation between the interdigitated chains; calculated $r$-to-$\alpha$ relation; schematic cartoons of the closely packed tilt angles; Gaussian fitting of the interactive energies on various nSAMs.

## Acknowledgement

The authors are grateful to the financial support of the Ministry of Science and Technology, Taiwan (MOST 110-2112-M-006-025-MY2 and NSTC 112-2628-E-131-001-MY4). We appreciate Dr. Chan Luo, Dr. Ye Huang, and Dr. Guillermo C. Bazan's kind help on the 2D-GIWAXS spectra when they were working in the UCSB. Use of the Stanford Synchrotron Radiation Lightsource, SLAC National Accelerator Laboratory, supported by the U.S. Department of Energy, Office of Science, Office of Basic Energy Sciences. We thank to Dr. Feng-Yin Chang's, Dr. Wang-Long Li's, and Dr. Jui-Chao Kuo's valuable comments and discussions. The kind support from Teledyne Princeton Instruments for the discontinued spectrometer is especially appreciated.